\def\kms{{\rm km\ s^{-1}}}
\def\etal{{\it et~al.\ }}
\def\msun{\rm M_{\odot}}
\def\arcsec{$^{\prime\prime}$}
\def\lya{Ly$\alpha$}

\documentstyle[psfig,12pt,koo_moriond]{article}
\begin{document}

\heading{DYNAMICS OF DISTANT NORMAL GALAXIES}

\author{David C. Koo $^{1}$} 
{$^{1}$ UCO/Lick Observatory, Department of
Astronomy and Astrophysics, \\
University of California, Santa Cruz, CA,
USA.}

\begin{moriondabstract}


Masses of galaxies are beginning to be measured systematically at
redshifts $z > 0.3$. Such data provide powerful and unique links and
clues to theories, simulations, and our understanding of early galaxy
formation and evolution. Discriminating masses of different kinds
(e.g., dark matter, old stars, HI gas) remains difficult, but a wide
variety of techniques for measuring total dynamical masses are being
explored and found practical. I highlight three recent optical studies of
dynamical masses: 1) the emission line velocity widths of blue
galaxies and rotation curves of spirals  that trace evolution in the
Tully-Fisher relation to $z \sim 1$; 2) the absorption line velocity
dispersions of early type galaxies in the field and clusters to track
evolution in the Fundamental Plane to $z
\sim 0.85$; and 3) the kinematics of  high redshift ($z \sim 3$)
Lyman-drop galaxies to constrain their nature, descendents, and
progenitors. The next decade shows promise of an explosive growth in
this infant field of measuring masses, especially in having much
larger sample sizes, higher precision and S/N, improved techniques and
tools, and probes of more diverse kinds of mass.

\end{moriondabstract}

\section{Introduction}

For over 20 years since the pioneering surveys of cluster galaxies by
Butcher and Oemler \cite{BO} and field galaxies by Peterson et
al. \cite{P79}, observations of distant normal galaxies have continually been
intensified to probe their formation and evolution. Such
explorations started with only photometric data in the form of number
counts and colors, as well as clustering statistics and photometric
redshifts. Spectroscopic redshifts and other low spectral resolution
information, such as star formation rates from [OII] line strengths or
age indices from the HK 4000 \AA \ break, soon followed. Even more
recently, studies have progressed to include morphologies, sizes, and
structures from HST images. The reader is referred to the contribution
in this volume by Hammer for an up-to-date overview of the global
properties of galaxy evolution.

In the last several years, a new and very powerful dimension has been
added to the suite of tools to explore distant galaxies, namely
internal kinematics and dynamics (masses). Although the potential was discussed
over a decade earlier by Kron \cite{K87}, only recently have advances
in instrumentation and access to larger telescopes allow such programs
to be practical. 

\subsection{Varieties of Mass}

Mass is a term whose exact meaning is often unspecified and dependent
upon the user and context. Confusion arises because mass, especially
in its baryonic forms, come in many
varieties. Dark matter, e.g.,  can be hot or cold,  while gas, stars, dust,
planets, and black holes can be further divided depending on its
location (bulge, disk, halo), time (young or old) or redshift (high or
low), color (blue or red) or temperature (hot or cold), different
elements or state (e.g., HI, HII, H$_2$, CO, gas, hot or warm dust),
etc.

 The importance of securing mass measurements for the study of distant
galaxy evolution appears obvious. After all, total or dynamical mass is
a fundamental physical property of galaxies.  And, given that star
formation remains a poorly understood physical process, mass possesses
much closer and more direct ties to our current theories and
simulations of galaxy formation than that from luminosity.  Moreover,
dynamical mass provides a new, very rich dimension to explore galaxy
formation and evolution that is in principle independent of luminosity
variations; that is conserved in isolated volumes; and which is likely
to be related to the bias in the clustering behavior of galaxies. Mass
can be measured for individual galaxies, to yield, e.g., M/L; to
assess whether an object is a bursting dwarf; to estimate the fraction
of dark matter; or to determine the relative proportions of stars and
gas. Due to the diversity of galaxy types and properties, however,
masses for statistically complete samples are more likely to reveal
convincing, unique, and well characterized evidence for
evolution. Such samples are essential to yield, e.g., the evolution in
the volume density of galaxies with different velocity widths or M/L,
or the clustering amplitude of galaxies as a function of mass or M/L.

Yet, despite their importance and potential, mass measurements for
distant galaxies have been sparse. In part this lack of data is due to
the greater difficulty in making such observations as compared to
redshifts alone. Also, many astronomers have the common perception
that mass and light are so tightly correlated for most local galaxies
that additional efforts to obtain mass measurements of faint distant
galaxies would be redundant and expensive.

\subsection{Techniques to Measure Mass}

For certain kinds of mass, the radiation itself is a direct measure.
Examples include HI from 21cm flux, old stars from rest-frame
near-infrared luminosities, or dust mass from submm if the temperature
is well constrained. Indirect measures include HI from CO, total
stellar mass from light assuming some average M/L, total stellar mass
from star formation rates and assumed lifetimes, or mass of gas from
column densities and sky coverage as probed by QSO absorption
lines. Instead of the masses of various subcomponents, the total mass
(dark or luminous) is almost always also desired. A relatively new
technique to measure total mass exploits gravitational lensing, which
can be applied to individual galaxies or clusters via strong lensing
events; to statistical samples of galaxies via galaxy-galaxy weak
lensing; or to the large scale mass power spectrum via global weak
lensing patterns.

For the vast bulk of galaxies, total masses are estimated from $v^2
r$, which assumes viralization and that the observed internal
kinematic velocities ($v$) and sizes ($r$) are reliable tracers of the
gravitational potential structure and depth. The kinematics of each
galaxy can be extracted via 1-D, 2-D, or 3-D of information.

In the 1-D case of flux versus wavelength, the {\it integrated}
spectrum of a galaxy is used to extract the velocity widths of
emission lines (as studied for HI gas via 21cm observations or for
ionized gas in star forming galaxies via optical spectra) or the
velocity dispersions of stellar absorption lines (e.g., as used in the
Faber-Jackson or Fundamental Plane relations for elliptical galaxies).
In general, 1-D optical data are obtained with a single aperture or
individual optical fiber for each galaxy.

In 2-D, the addition of spatial resolution along one axis provides
rotation curves. The extra dimension yields qualitatively new types of
information, such as the radial change in M/L that can reveal the
amount of dark matter; the velocity distortions that might be
signatures of mergers; or gradients in the velocity dispersion of
disks that may reflect the rate of heating by satellites. Most of
these measurements are based on long-slit apertures, usually along the
major axis of a galaxy.

In 3-D, the remaining angular spatial dimension is added and provides
information on the position angle, inclination, possible asymmetries,
etc. in the kinematics of a galaxy \cite{B97}. The most common optical
instruments to extract such 3-D data for distant galaxies include
integral field units with bundles of optical fibers, Fabry-Perot
systems, and ramped narrow-band filter systems.

\section{Overview of Kinematic Studies of Distant Galaxies}

Given the difficulty of securing just the redshift, much less any line
profile information, distant galaxy spectral surveys of the 1980's
through the early 1990's relied on relatively low spectral resolution
spectrographs, typically with velocity resolutions of $\sim$ 500 to
1000 $\kms$, which is inadequate for internal kinematic studies of
most galaxies. Although 4-m to 6-m class telescopes are able to yield
kinematics for brighter galaxies at intermediate redshifts, the vast
bulk of distant galaxy kinematic observations has come from the Keck 10-m
Telescopes that started operating in 1994. The following subsections
highlight the major scientific results from these pioneering efforts.

\subsection{Emission Line  Kinematics}

Unless distant galaxies produce significant gas motions unrelated to
their mass or their emission line spatial distribution is
unrepresentative of the underlying structure, the widths of emission
lines from their integrated spectra should provide a good diagnostic
of their gravitational potential.  Motivated by three factors: 1) the
relatively good resolution of the Low Resolution Imaging Spectrograph
(LRIS \cite{O95}) of less than 100 $\kms$; 2) the availability of high
spatial resolution images from the Hubble Space Telescope to derive
inclination angles and sizes of galaxies; and 3) the desire to test a
number of claims for dramatic evolution of moderate redshift field
galaxies, the Deep Extragalactic Evolutionary Probe (DEEP: see URL:
http://www.ucolick,.org/$^{\sim}$deep/home.html) team initiated a
number of pilot surveys of internal kinematics \cite{K98}. One early
LRIS survey included 18 apparent spirals brighter than $I
\sim 22$. They were found to have redshifts $z \sim $ 0.2 to 0.8 and
integrated velocity widths that yielded M/L ratios only about one
magnitude brighter than that of local spirals
\cite{F96}. Due to LRIS being down at the last
moment of a scheduled run, another DEEP survey used the only available
instrument, the High Resolution Spectrograph
(HIRES: \cite{V94}) with a velocity resolution of less than 10 $\kms$ \
on Keck,  to observe a sample of 17 very blue, very luminous ($\sim
L^*$) compact galaxies with $B \sim 20$ to 23 and redshifts $z
\sim 0.1$ to 0.7 \cite{K95}.  The surprisingly small velocity dispersions
of 30 to 60 $\kms$ \ for most of these galaxies, along with their small sizes,
suggested that they were similar to HII galaxies and perhaps the
progenitors of luminous spheroidal galaxies seen locally
\cite{G96}\cite{G98}. Similar results were found for a sample of 6
very strong emission-line galaxies in cluster Cl 0024+1654 at $z \sim
0.4$ \cite{K97}. Further LRIS observations of even fainter ($I <
23.5$) compact galaxies found in the flanking fields of the Hubble
Deep Field showed that such galaxies are likely to be significant
contributors to the star formation rate at high redshifts $z \sim 0.8$
and that the galaxies with high specific star formation rates (per
mass) may be shifting to higher masses at higher redshifts
\cite{P97}\cite{G97}.

In the meantime, two other groups pushed the limits of 4-m class
telescopes with instruments that yielded velocity resolutions of about
50 $\kms $. One group used the AUTOFIB fiber-optics spectrograph on
the AAT to explore the kinematics of 24 blue galaxies with $B \sim
21.5$ at redshifts $z \sim 0.25$\cite{R97}. The other group used the
Subarcsec Imaging Spectrograph (SIS: \cite{LeF94}) on the CFHT to
examine 24 others with $I < 22$ and with $z \sim 0.6$ \cite{M99}. Both
groups claimed detection of 1 mag to 2 mag of luminosity brightening
compared to local counterparts. As previously found \cite{G97}, the
evidence for evolution was particularly strong among compact galaxies
\cite{M99}.  These surveys reveal that the masses of otherwise
similarly luminous galaxies (typically near $L^*$) span from genuine
dwarfs with $10^9 \msun$ to bonafide giants with well over $10^{11}
\msun $. The poor correlation of mass with luminosity found for some,
albeit very blue, galaxies justifies the clear need for internal
kinematics measurements in the study of distant galaxy evolution.

\begin{figure}
\centerline{\vbox{\psfig{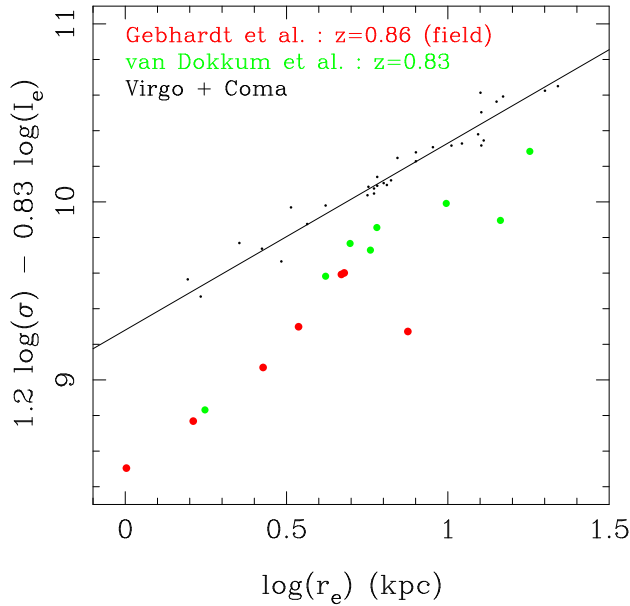}}}

\caption[]{
The Fundamental Plane for a sample of cluster \cite{Van98} and field
\cite{G99} early-type galaxies at redshift $z \sim 0.85$. The vertical
offset in the surface brightness ($I_e$) corresponds roughly to a
magnitude of brightening. The reality of the differences in the
vertical offset from the small to large galaxies is tantalizing
evidence for possible differences in their IMF or formation history,
but remains uncertain.  This figure and pre-publication data were
kindly provided by K. Gebhardt.  

}
\end{figure}

Finally, we note that even the internal kinematics of the important,
but very faint ($R \sim$ 24 to 26 mag), population of Lyman-drop galaxies
\footnote {Lyman-drop is preferred over the more popular term of
Lyman-break galaxies, since the \lya \ forest also contributes to the
shape of the spectrum.} has been studied via velocity widths of their
emission lines. A major issue is whether the masses of these galaxies
are large, as inferred from theory and their observed strong
clustering properties \cite{St98}, or are low if they are instead
star-bursting pre-merger components, starbursting subcomponents within
larger-mass halos, or progenitors of genuine dwarf galaxies \cite{L97}
\cite{S99}. The spectral evidence is meager but favors small masses, with
emission line widths of generally less than $100 \kms$ \ from \lya \ as
measured in the optical \cite{L97} or from the more reliable
$H_{\beta}$ and [OIII] lines that were obtained in the near-infrared
\cite{P98}.

\subsection{Absorption Line Velocity Dispersions}

These measures are considerably more difficult than the widths of
strong emission lines, due to the need to have high S/N in the
continuum. Thus far, absorption line widths have mainly been measured
in distant, luminous early-type galaxies. This challenging work
typically requires the use of 8-10m class telescopes, moderately good
spectral resolution, and a corresponding set of stellar templates for
cross-correlations. The handful of studies compare the M/L ratio and Fundamental Plane of distant versus local
E/S0 for rich clusters of galaxies
\cite{Van96}\cite{Van98}\cite{Kelson97}\cite{Z97} and field galaxies
\cite{G99}. The results are reassuringly consistent, with any changes
being a good match to just mild passive luminosity evolution, which 
predict $\sim 1$ mag of brightening by redshifts $z \sim$ 0.85
(see Fig. 1).

\begin{figure}
\centerline{\vbox{\psfig{figure=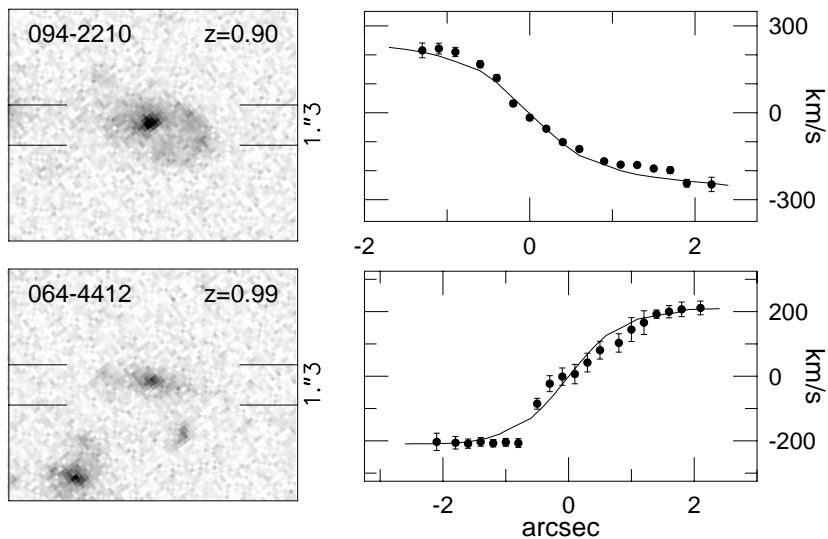,height=7.cm}}}

\caption[]{
Examples of the rotation curves measured for 
two high redshift galaxies, the upper with total $I \sim 21.4$ and the lower
with $I \sim 22.4$ \cite{Vogt97}.
}
\end{figure}

\begin{figure}
\centerline{\vbox{\psfig{figure=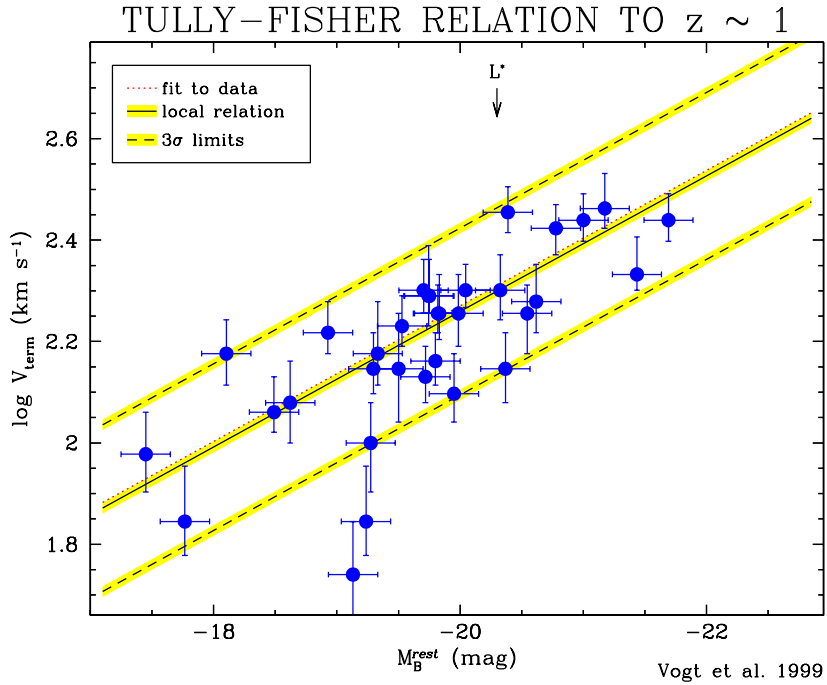,height=9.cm}}}

\caption[]{
The Tully-Fisher relation from DEEP for faint, distant spiral galaxies
up to redshifts $z \sim 1$ \cite{Vogt99}. Note the consistency with
the local TF relation. This figure and pre-publication data were
kindly provided by N. Vogt.
}
\end{figure}

\begin{figure}
\centerline{\vbox{\psfig{figure=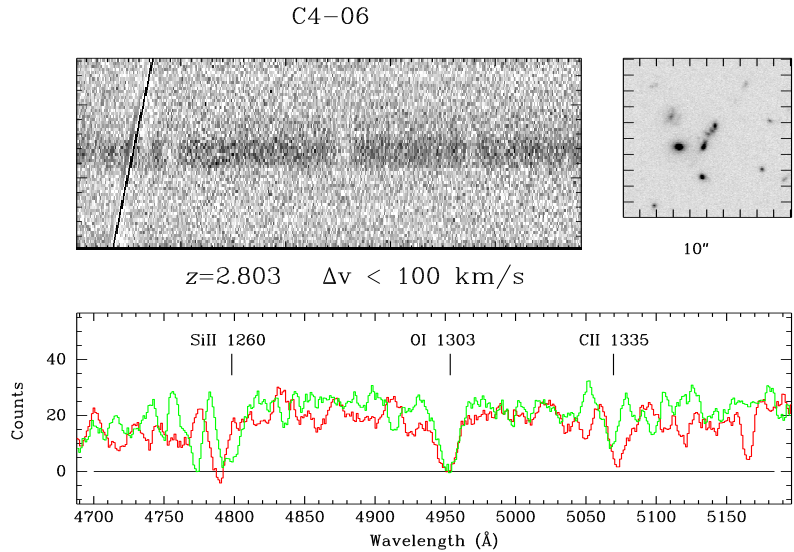,height=8.5cm}}}

\caption[]{
Upper left panel shows the 2-D sky-subtracted spectra for the very
extended, or possibly two-component, Lyman-drop galaxy shown in the
center of the upper right panel. The bottom figure shows part of the
spectrum of each component separately. A cross-correlation of the
spectra yields no statistically detectable velocity shift.  Adoption
of an upper limit of 100 $\kms$ for the differential velocity, no
inclination corrections, and the apparent angular separation as a
measure of the actual size of the system yields an upper limit to the
mass of $10^{10}\msun$. This source, C4-06, was originally discovered
in the HDF by Steidel \etal with Keck \cite{St96} and reobserved at
higher spectral resolution to derive the kinematics
\cite{L98}. This figure was kindly provided by J. Lowenthal. 
}
\end{figure}

\subsection{Spatially Resolved Kinematics}

These observations require both good spectral resolution and
moderately high spatial resolution in the spectra, as well as
information on inclination and position angle of the major axis for
each galaxy. Simulations are generally needed to derive such parameters
as the terminal velocity, by accounting  for seeing, slit width, optical
PSF, inclination, etc.

At the lower redshift regime, 4-m to 5-m class programs have already
yielded interesting, but apparently inconsistent, results. In one
case, the claim is that 40 high-quality rotation curves of spirals
with redshifts $z \sim 0.2$ to 0.4 show {\it no evidence for
evolution} in the Tully-Fisher relationship between luminosity and
velocities {\it after} corrections are made for the colors of the
galaxies \cite{B97} \cite{B98a} \cite{B98b}
\cite{B98c}. Another group studied a sample of 22 strong [OII]
emitting galaxies with $z \sim 0.35$ and $R \sim 21$ with the SIS and
Multi Object Spectrograph (MOS: \cite{LeF94}) on CFHT \cite{S98}. They
find nearly 2 mag of luminosity brightening when compared to local
samples. At yet higher redshifts to $z \sim 1$, the Keck Telescope
with LRIS has yielded over 30 decent rotation curves for galaxies that
appear to be relatively large spirals in HST images \cite{Vogt96}
\cite{Vogt97} \cite{Vogt99} (see Fig. 2 for examples). As seen in
Fig. 3, there is little, if any, offset ($< 0.5$ mag) between this
high-redshift sample and that of local Tully-Fisher samples. This
result is particularly interesting for two reasons. First, at least
some theories predict quite rapid disk evolution \cite{M98}, while
these observations, along with the evidence for little change in the
volume density of large disk galaxies \cite{Lilly98}, appear to
suggest otherwise. Second, the stark contrast of the spiral results to
the evidence for much stronger evolution among compact galaxies
(subsection 2.1) is suggestive of an important, missing component in
our understanding of galaxy evolution.

Although Lyman-drop galaxies are generally too small ($\sim
$0.2\arcsec) to yield rotation curves with present ground-based
systems, some appear to have multiple or extended components, which
then allow spatially resolved kinematics to be gathered. So far,
results exist only for two galaxies, with one consistent with a very
low mass of less than 10$^{10} \msun $ (see Fig. 4), while the other
has a more typical mass of 3 10$^{11}\msun $ for luminous
galaxies \cite{L98}. Given the unknown inclination of the plane of their relative
motions and their true sizes or separations, more data are clearly
needed to obtain the needed statistics to assess whether Lyman-drop
galaxies are generally large or small mass systems.

\section{Future Prospects}

The aforementioned pioneering surveys clearly demonstrate the
feasibility of acquiring kinematics and dynamics of high redshift
galaxies. The resultant scientific conclusions, even if currently
tentative and based on limited samples, highlight the potential to
extract critical and unique clues to their nature and evolution.  The
whole field of observing the dynamics of distant galaxies is still in
its infancy and will soon experience an explosive growth.

\begin{figure}
\centerline{\vbox{\psfig{figure=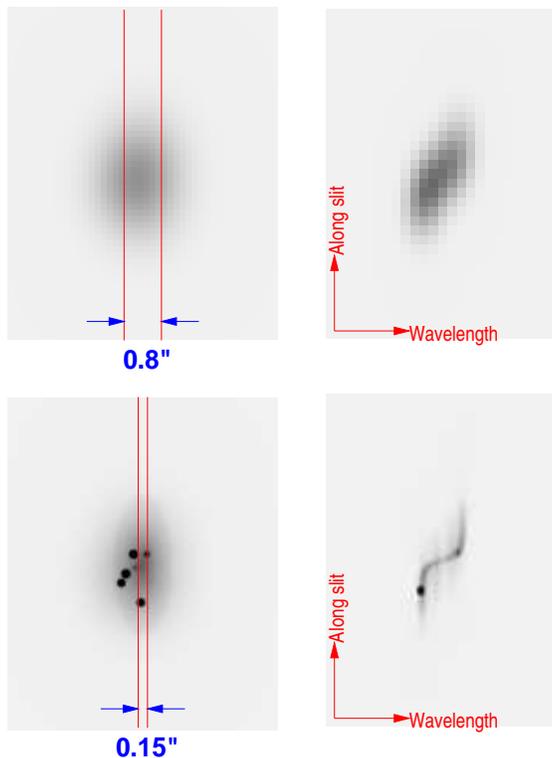,height=10.cm}}}

\caption[]{
Simulations of an emission line rotation curve with and without using
adaptive optics on the Keck 10-m Telescope. The target has an
exponential disk with a scale length of 0.2\arcsec, inclination of
45$^{\circ}$, and a maximum velocity in the rotation curve of 40
$\kms$. The upper left panel shows the assumed placement of a
0.8\arcsec \ slit over the image under seeing of 0.6\arcsec . The
upper right panel shows the expected emission line using the DEIMOS
instrument \cite{D98} with 0.12\arcsec \ pixels in the vertical
direction and 0.32 \AA \ pixels in wavelength. The lower two panels
show the vast improvement in both imaging and spectroscopy when the
adaptive optics system is assumed to have a Strehl of 0.33 with a
0.05\arcsec \ core and a much narrower slit of 0.15\arcsec \ is used
with a hypothetical spectrograph having 0.03\arcsec \ pixels in the
vertical direction and a spectral resolution of 0.08 \AA \ per
pixel. This simulation was produced and generously provided by A. C.
Phillips.  
}
\end{figure}

Over the next decade, we will see the completion of several major
redshift surveys with high enough spectral resolution to yield
internal kinematics. These include the Sloan Digital Sky Survey
(SDSS:\cite{Lov98}), which will firmly set the foundation for the
local internal kinematic properties of diverse galaxy types; the next phase of
DEEP, which will aim for about 50,000 galaxies at $z\sim 0.9$
\cite{D98}; and the VIRMOS survey on the VLT \cite{LeF98}, which will
access not only the optical but also the near-IR that reaches
the rest-frame optical for redshifts beyond $z \sim 1$.  In principle, such
kinematic surveys can be used to estimate the masses of galaxies on
larger scales than actually observed in the luminous portions of
galaxies, by measuring the distribution of relative velocities of
galaxy pairs at different separations. More direct measures of
dynamical mass on larger scales are likely to come from weak lensing
surveys, several of which are already underway; some will take
advantage of photometric redshifts to vastly improve the S/N by
discriminating between foreground and background sources. Moreover,
adaptive optics (see Fig. 5) along with NGST will yield detailed
kinematics of very high redshift and even the most compact
galaxies. Finally, planned enhancements to submm, mm, and radio
telescopes should provide direct measures of the amount and motions of
various forms of gas in distant galaxies (a recent example at $z \sim 3.4$ uses 21cm in
absorption \cite{Br97}).

The science from these large samples of mass measurements for distant
galaxies will be revolutionary. Galaxy formation, e.g., will be
observed as a physical process involving the hierarchical buildup of
mass, rather than the current popular trend of concentrating on galaxy
formation as reflected by the star formation rate (SFR), i.e. just the
conversion of gas into stars.  One diagnostic to explore this merging
process includes the evolution of the comoving volume density function
(EOCVDF) of mass or velocity; another is to combine the information
with luminosity to measure the EOCVDF(M/L). With the advent of high
spatial resolution spectroscopy that enables separation of the structure and
kinematics of galaxy {\it subcomponents}, the EOCVDF(disk mass, M/L,
scale-length) or EOCVDF(bulge mass, M/L, size) will be dramatic
enhancements to current studies of the Tully-Fisher and Fundamental
Plane relations. Adding other dimensions of
information such as SFR, chemical abundances, and stellar ages will
only add richness and depth to the forthcoming revolution in distant
galaxy studies.  As emphasized in the introduction, mass comes in
diverse forms, so the previous discussion applies also for dark
matter, halo, HI, etc. as new telescopes and tools allow
discrimination among them.

This revolution will extend even to cosmology. Large-scale structure
and bias, e.g., will be characterized and discriminated by mass (mass
power spectrum) instead of light. Independent tests of the global
geometry can be undertaken by using the Fundamental Plane of
ellipticals \cite{Bender98} or via the classical volume tests
\cite{L88}. The latter is feasible through the EOCVD of quantitities
that are expected to be conserved, such as the total baryonic mass
(obtained by combining results from X-ray to radio) or the total mass
(from gravitational lensing).

\begin{acknowledgements}{ DEEP was initiated by the Berkeley Center for
Particle Astrophysics (CfPA) and has been supported by various NSF,
NASA, and STScI grants over the years, including NSF AST-9529098 and
STScI AR-07532.01-96.  K. Gebhardt, N. Vogt, A. C. Phillips, and
J. Lowenthal are especially thanked for providing the figures. I also
thank V. Rubin and R. Kron for their encouragement to me in the early 1980's
to explore the dynamics of distant galaxies.
}
\end{acknowledgements}


%
\vfill
\end{document}